\documentclass[%
reprint,
 superscriptaddress,
 amsmath,amssymb,
 aps,
prb,
]{revtex4-1}

\usepackage{graphicx}
\usepackage{dcolumn}
\usepackage{bm}

\usepackage[ 
margin=0.75in,
]{geometry}

\def\v2o3{V$_2$O$_3$}
\def\rno{$RE$NiO$_{3}$}
\def\musr{$\mu$SR}

\def\nno{NdNiO$_3$}
\def\pno{PrNiO$_3$}
\def\snno{Sm$_{0.75}$Nd$_{0.25}$NiO$_3$}
\def\nlnost{Nd$_{0.7}$La$_{0.3}$NiO$_3$}
\def\nlnosf{Nd$_{0.6}$La$_{0.4}$NiO$_3$}
\def\nlnoff{Nd$_{0.5}$La$_{0.5}$NiO$_3$}
\usepackage[usenames]{color}
\definecolor{gray}{gray}{0.5}

\begin{document}

\preprint{}

\title{
Volume-wise destruction of the antiferromagnetic Mott insulating state through quantum tuning
}

\author{Benjamin A. Frandsen}
\affiliation{%
 Department of Physics, Columbia University, New York, NY 10027, USA.
}%

\author{Lian Liu}
\affiliation{%
 Department of Physics, Columbia University, New York, NY 10027, USA.
}%

\author{Sky C. Cheung}
\affiliation{%
 Department of Physics, Columbia University, New York, NY 10027, USA.
}%

\author{Zurab Guguchia}
\affiliation{%
 Laboratory for Muon Spin Spectroscopy, Paul Scherrer Institute, CH-5232 Villigen PSI, Switzerland
}%

\author{Rustem Khasanov}
\affiliation{%
	Laboratory for Muon Spin Spectroscopy, Paul Scherrer Institute, CH-5232 Villigen PSI, Switzerland
}%

\author{Elvezio Morenzoni}
\affiliation{%
 Laboratory for Muon Spin Spectroscopy, Paul Scherrer Institute, CH-5232 Villigen PSI, Switzerland
}%

\author{Timothy J. S. Munsie}
\affiliation{%
 Department of Physics and Astronomy, McMaster University, Hamilton, Ontario L8S 4M1, Canada.
}%

\author{Alannah M. Hallas}
\affiliation{%
 Department of Physics and Astronomy, McMaster University, Hamilton, Ontario L8S 4M1, Canada.
}%

\author{Murray N. Wilson}
\affiliation{%
 Department of Physics and Astronomy, McMaster University, Hamilton, Ontario L8S 4M1, Canada.
}%

\author{Yipeng Cai}
\affiliation{%
 Department of Physics and Astronomy, McMaster University, Hamilton, Ontario L8S 4M1, Canada.
}%

\author{Graeme M. Luke}
\affiliation{%
 Department of Physics and Astronomy, McMaster University, Hamilton, Ontario L8S 4M1, Canada.
}%
\affiliation{%
 Canadian Institute for Advanced Research, Toronto, Ontario L8S 4M1, Canada.
}%

\author{Bijuan Chen}
\affiliation{ %
Institute of Physics, Chinese Academy of Sciences, Beijing, China.
} %

\author{Wenmin Li}
\affiliation{ %
Institute of Physics, Chinese Academy of Sciences, Beijing, China.
} %

\author{Changqing Jin}
\affiliation{ %
Institute of Physics, Chinese Academy of Sciences, Beijing, China.
} %

\author{Cui Ding}
\affiliation{ %
Department of Physics, Zhejiang University, Hangzhou, China.
} %

\author{Shengli Guo}
\affiliation{ %
Department of Physics, Zhejiang University, Hangzhou, China.
} %

\author{Fanlong Ning}
\affiliation{ %
Department of Physics, Zhejiang University, Hangzhou, China.
} %

\author{Takashi U. Ito}
\affiliation{ %
	Advanced Science Research Center, Japan Atomic Energy Agency, Tokai, Ibaraki 319-1195, Japan.
} %

\author{Wataru Higemoto}
\affiliation{ %
	Advanced Science Research Center, Japan Atomic Energy Agency, Tokai, Ibaraki 319-1195, Japan.
} %

\author{Simon J. L. Billinge}
\affiliation{%
 Department of Applied Physics and Applied Mathematics, Columbia University, New York, NY 10027, USA.
}%
\affiliation{%
Condensed Matter Physics and Materials Science Department, Brookhaven
National Laboratory, Upton, New York 11973, USA.
}%

\author{Shoya Sakamoto}
\affiliation{ %
	Department of Physics, University of Tokyo, Bunkyo-ku, Tokyo 113-0033, Japan.
} %

\author{Atsushi Fujimori}
\affiliation{ %
	Department of Physics, University of Tokyo, Bunkyo-ku, Tokyo 113-0033, Japan.
} %

\author{Taito Murakami}
\affiliation{ %
Department of Energy and Hydrocarbon Chemistry, Graduate School of Engineering, Kyoto University, Nishikyo, Kyoto 615-8510, Japan.
} %

\author{Hiroshi Kageyama}
\affiliation{ %
Department of Energy and Hydrocarbon Chemistry, Graduate School of Engineering, Kyoto University, Nishikyo, Kyoto 615-8510, Japan.
} %

\author{Jose Antonio Alonso}
\affiliation{ %
	Instituto de Ciencia de Materiales de Madrid (ICMM), CSIC, Madrid, Spain.
} %

\author{Gabriel Kotliar}
\affiliation{ %
Department of Physics \& Astronomy, Rutgers University, Piscataway, New Jersey 08854-8019, USA.
} %

\author{Masatoshi Imada}
\affiliation{ %
Department of Applied Physics, University of Tokyo, 7-3-1 Hongo, Tokyo, 113-8656, Japan
} %

\author{Yasutomo J. Uemura}
\thanks{Author to whom correspondence should be addressed: Y.J.U., tomo@lorentz.phys.columbia.edu}
\affiliation{%
 Department of Physics, Columbia University, New York, NY 10027, USA.
}%

\date{\today}

\maketitle

\textbf{Metal-to-insulator transitions (MITs) are a dramatic manifestation of strong electron correlations in solids~\cite{imada;rmp98}. The insulating phase can often be suppressed by quantum tuning, i.e. varying a nonthermal parameter such as chemical composition or pressure, resulting in a zero-temperature quantum phase transition (QPT) to a metallic state driven by quantum fluctuations, in contrast to conventional phase transitions driven by thermal fluctuations~\cite{vojta;rpp03}. Theories of exotic phenomena known to occur near the Mott QPT such as quantum criticality and high-temperature superconductivity~\cite{keime;n15} often assume a second-order QPT, but direct experimental evidence for either first- or second-order behavior at the magnetic QPT associated with the Mott transition has been scarce and further masked by the superconducting phase in unconventional superconductors. Most measurements of QPTs have been performed by volume-integrated probes, such as neutron scattering, magnetization, and transport, in which discontinuous behavior, phase separation, and spatially inhomogeneous responses are averaged and smeared out, leading at times to misidentification as continuous second-order transitions. Here, we demonstrate through muon spin relaxation/rotation (\musr) experiments on two archetypal Mott insulating systems, composition-tuned \rno\ ($RE$=rare earth element) and pressured-tuned \v2o3, that the QPT from antiferromagnetic insulator to paramagnetic metal is first-order: the magnetically ordered volume fraction decreases to zero at the QPT, resulting in a broad region of intrinsic phase separation, while the ordered magnetic moment retains its full value across the phase diagram until it is suddenly destroyed at the QPT. As an unambiguous demonstration of a first-order magnetic QPT in non-superconducting Mott systems, these findings bring to light a surprising universality of the pressure-driven Mott transition in three spatial dimensions, revealing the importance of phase separation in a significant region of parameter space near the QPT and calling for further investigation into the role of inelastic soft modes and the nature of dynamic spin and charge fluctuations underlying the transition.} 

\begin{figure}
	\includegraphics[width=80mm]{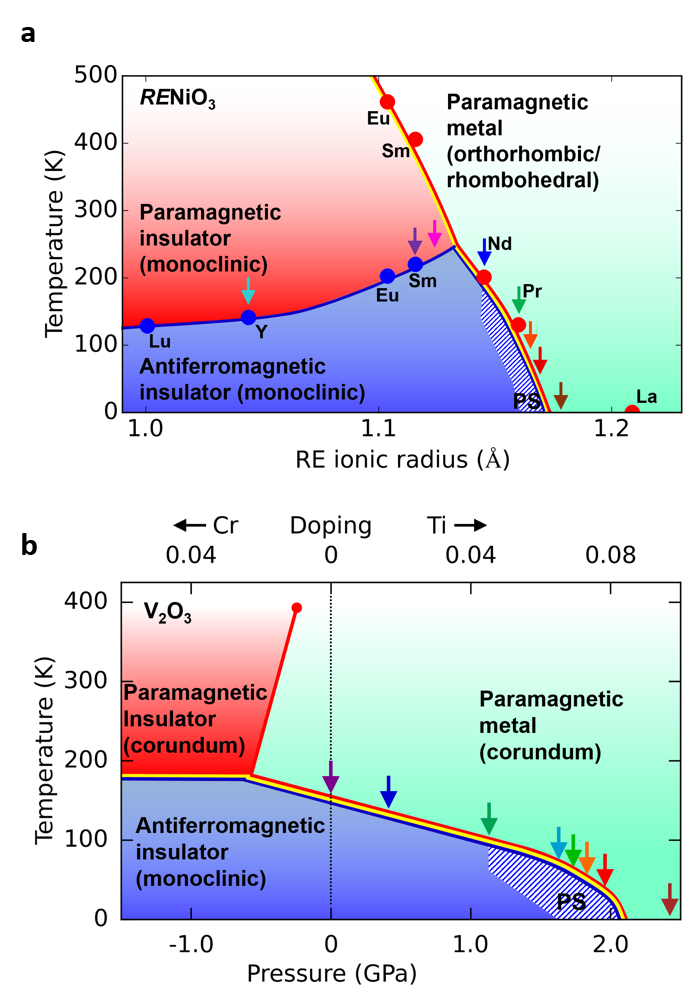}
	\caption{\label{fig:phasediagram} \textsl{Phase diagrams for canonical Mott systems \rno\ and \v2o3. (a) \rno\ phase diagram, with temperature along the vertical axis and rare-earth ionic radius along the horizontal axis. The red curve indicates a metal-insulator transition on cooling, blue a paramagnet-antiferromagnetic transition, and yellow a structural transition. The colored circles represent phase boundaries for the stoichiometric compounds determined by previous studies. Colored arrows indicate compositions studied in the current work. The QPT occurs at a radius of approximately 1.175 \AA. ``PS'' = phase separated. After Ref.~\onlinecite{torra;prb92}. (b) \v2o3\ phase diagram, with temperature along the vertical axis and hydrostatic pressure along the horizontal axis. The QPT occurs at a pressure of approximately 2.0~GPa. Doping with Ti and Cr is shown on the upper horizontal axis for comparison. All colored curves and symbols are the same as in (a). After Ref.~\onlinecite{mcwha;prb73}.}}
	
\end{figure}

The low-temperature antiferromagnetic Mott insulating state (defined here broadly as a correlation-induced insulating state) in \rno\ and \v2o3\ can be tuned systematically by straightforward methods: variation of the rare-earth ion~\cite{torra;prb92} in \rno\ and application of hydrostatic pressure~\cite{mcwha;prl69} in stoichiometric \v2o3, making them ideal systems for studying correlation-driven MITs. \rno\ possesses a distorted perovskite structure in which the average rare-earth ionic size can be continuously controlled by solid solution of different rare earths~\cite{torra;prb92}, effectively applying a chemical pressure and producing the temperature---ionic-radius phase diagram shown in Fig.~\ref{fig:phasediagram}a. An abrupt thermal phase transition (solid red curve) from a high-temperature metal to a low-temperature insulator takes place at a temperature that depends strongly on ionic radius, with a paramagnetic-antiferromagnetic transition (blue curve) occuring simultaneously for some compounds and at lower temperature for others. A structural response (yellow curve) is also observed when cooling through the MIT: the unit cell volume increases slightly, the rotational distortion of the NiO$_6$ octahedra is enhanced, and the crystallographic symmetry is lowered from orthorhombic to monoclinic~\cite{alons;prl99}. As the ionic radius increases, the magnitude of this distortion lessens in both the metallic and insulating phases, and the temperature of the thermal MIT decreases. At the estimated critical radius of $\sim1.175$~\AA, the MIT temperature is suppressed to zero, resulting in a QPT from an antiferromagnetic insulator to a paramagnetic metal. Varying the rare-earth ion leaves the electron count unchanged but alters the width of the relevant electron energy bands by changing the crystal structure; hence, this constitutes a bandwidth-controlled Mott QPT~\cite{imada;rmp98}.

The exact mechanism of the MIT remains a topic of debate even after multiple decades. Various scenarios have been proposed for the origin of the insulating state, including the opening of a charge-transfer gap~\cite{torra;prb92}, orbital ordering~\cite{garci;epl92}, and charge ordering~\cite{alons;prl99,garci;prb09}, but certain experimental observations have been difficult to reconcile with each of these scenarios~\cite{abbat;prb02,loren;prb05,garci;prb09,boden;jpcm11}. Recent progress has led to a proposed site-selective Mott transition~\cite{park;prl12,johns;prl14} that successfully explains many of the unusual electronic properties of \rno.

In the case of \v2o3, the application of hydrostatic pressure monotonically decreases the MIT temperature from $\sim$160~K at ambient pressure to 0~K at $\sim$2.0~GPa (see phase diagram in Fig.~\ref{fig:phasediagram}b), again resulting in a bandwidth-controlled QPT. The MIT is likewise accompanied by a volume increase and symmetry lowering (rhombohedral to monoclinic) in the insulating state. Replacing small amounts of V with either Ti or Cr produces similar effects as pressure~\cite{mcwha;prb73}, as shown on the upper horizontal axis of Fig.~\ref{fig:phasediagram}b. Doping with Cr acts as ``negative pressure'' and leads to a first-order transition at high temperature between a paramagnetic metal and paramagnetic insulator, dividing the overall \v2o3\ phase diagram into three parts, similar to \rno. However, the metallic phase produced by Ti doping orders magnetically~\cite{imada;rmp98} and is therefore qualitatively different from the pressure-induced metal, so these two tuning methods should not be viewed as entirely equivalent.


The MIT in \v2o3\ was classified early on as a Mott-Hubbard transition~\cite{mcwha;prl69}. Experimental and theoretical developments in the 1990s and 2000s, particularly from numerical techniques such as dynamical mean field theory (DMFT), extended the understanding of this system in terms of multiband, $S=1$ models, with consideration given to the interplay of spin and orbital degrees of freedom~\cite{rozen;prl95,kelle;prb04}. A detailed understanding of the subtle differences between the temperature-, doping-, and pressure-induced MITs remains a topic of active study~\cite{hansm;pss13}.

Most of the previous work on these materials has focused on the thermal MIT. The QPT along the horizontal axis of the phase diagram has received less attention, despite being vital for a full understanding of these materials and the pressure-driven Mott transition in general. The order of the QPT is significant. The critical point of a second-order QPT is called a quantum critical point (QCP) and forms the vertex of a fan-shaped ``quantum critical'' region in the phase diagram, which may be associated with unusual properties and novel electronic phases. For a first-order QPT, the system would be expected to exhibit more typically first-order behavior such as phase coexistence and abrupt changes in the ground state, not necessarily manifesting quantum criticality in the same way. Early numerical calculations~\cite{watan;jpsj04,misaw;prb07}, scaling analysis~\cite{imada;prb05}, and DMFT studies~\cite{chitr;prl99} suggested first-order behavior at the Mott QPT, but conclusive experiments on \rno\ and \v2o3\ have been lacking. Among the relatively few experiments related to the QPT in these materials are a neutron study~\cite{kadow;prl08} suggesting quantum critical spin fluctuations in heavily Ti-doped \v2o3, and very recent scanning tunneling spectroscopy measurements of thin-film NdNiO$_3$ and LaNiO$_3$ that can be interpreted as evidence for an underlying QCP~\cite{allen;arxiv14}. However, much remains unknown about the QPT in both of these systems, in part because volume-integrating probes like transport and neutron scattering are typically unable to distinguish between second-order and first-order transitions with phase separation.

To investigate this problem, we have utilized muon spin relaxation/rotation (\musr) to study several compositions of \rno, with rare earths indicated by the colored arrows in Fig.~\ref{fig:phasediagram}a: Y (cyan), Sm (purple), $Sm_{0.75}Nd_{0.25}$ (pink), Nd (blue), Pr (green), $Nd_{0.7}La_{0.3}$ (orange), $Nd_{0.6}La_{0.4}$ (red), and $Nd_{0.5}La_{0.5}$ (brown). We also studied \v2o3\ at several pressures spanning the phase diagram, again indicated by colored arrows in Fig.~\ref{fig:phasediagram}b: ambient pressure, 0.4~GPa, 1.29~GPa, 1.64~GPa, 1.73~GPa, 1.83~GPa, 1.96~GPa, and 2.43~GPa. The \musr\ technique exploits the asymmetric decay of muons into positrons to act as a highly sensitive probe of local magnetism. Details are provided in the Methods section. \musr\ experiments in zero externally applied magnetic field (ZF) and a weak transverse field (wTF) applied perpendicular to the initial muon spin direction were performed for each sample on a temperature grid spanning the thermal transition, providing a detailed picture of the phase diagram near the QPT. Importantly, \musr\ can measure the local order parameter and the magnetically ordered volume fraction independently from one another, and is therefore ideally suited to determine first- or second-order behavior at the QPT.

\begin{figure*}
	\includegraphics[width=140mm]{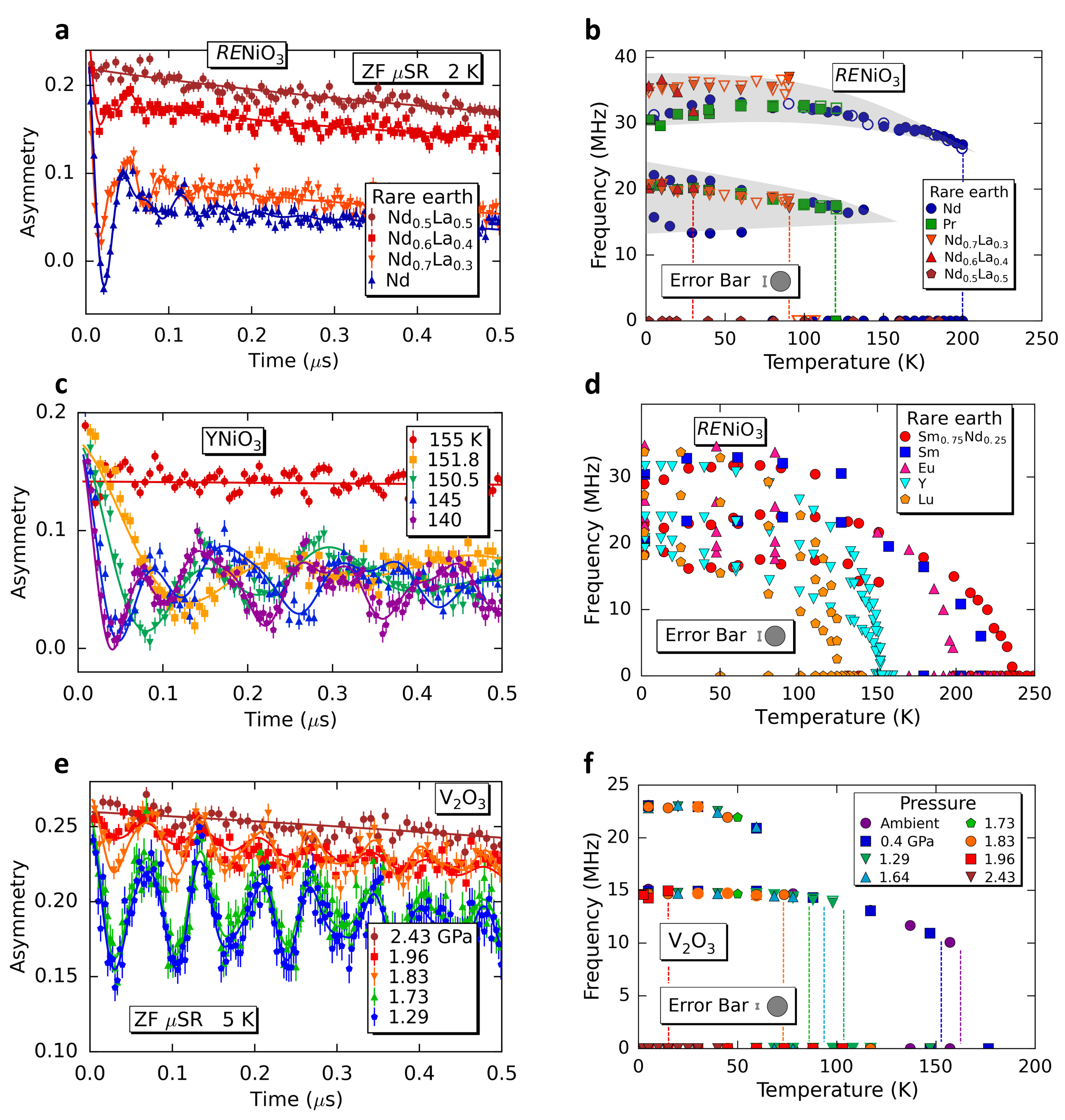}
	\caption{\label{fig:ZF} \linespread{1.1}\fontsize{10}{12}\selectfont{}\textsl{Results of ZF \musr\ experiments on \rno\ and \v2o3. (a) ZF time spectra taken at 2~K for four compounds of \rno\ near the QPT. The colored dots represent the data, the solid curves the fits. The three magnetically ordered compounds show nearly identical oscillation frequencies (hence identical moment sizes) but very different oscillation amplitudes (hence different ordered volume fractions). (b) Temperature dependence of the oscillation frequencies  for \rno\ compounds with first-order thermal phase transitions. Filled (open) circles represent data taken in a cooling (warming) sequence. All magnetically ordered compounds have two or three frequencies lying along two common bands (shaded gray regions), indicating that the ordered moment size does not change along the horizontal axis of the phase diagram. The large gray circle with the neighboring vertical bar indicates the typical estimated standard deviation (ESD) of the refined frequency compared to the symbol size. The colored dashed lines are guides to the eye showing the approximate transition temperature for each composition. (c) ZF spectra for YNiO$_3$ taken at various temperatures through the thermal phase transition. (d) Temperature dependence of the oscillation frequencies for \rno\ compounds with second-order-like thermal phase transitions, revealing the continuous development of the ordered moment size. (e),(f) Plots for pressure-tuned \v2o3\ corresponding to (a) and (b).}}
	
\end{figure*}

Representative ZF time spectra taken at 2~K on four compositions of \rno\ near the QPT are displayed in Fig.~\ref{fig:ZF}a. The spectrum for \nlnoff\ exhibits only slow relaxation with no oscillations, indicating an absence of magnetic order, as expected for this composition. In contrast, damped but coherent oscillations are seen for \nlnosf, placing this composition just to the left of the QPT. \nlnost\ and \nno\ likewise show oscillations indicative of magnetic ordering. 

For a given material, the ZF oscillation frequency and amplitude are directly proportional to the ordered moment size and the magnetically ordered volume fraction of the sample, respectively. From Fig.~\ref{fig:ZF}a, one clearly observes that the three ordered compounds have approximately the same oscillation frequency (hence moment size), but different oscillation amplitudes (hence OVFs). To better illustrate the frequency behavior, we plot in Fig.~\ref{fig:ZF}b the temperature dependence of the oscillation frequency for the compounds closest to the QPT. These frequencies were extracted from refinements described in the Methods section. The most striking feature of these results is that the two or three oscillation frequencies observed in each of the ordered compounds all lie along the same two frequency bands, illustrated by the shaded gray regions in Fig.~\ref{fig:ZF}b. This indicates that the saturated moment size is not changed by compositional tuning toward the QPT; rather, the full moment is destroyed abruptly and discontinuously at the QPT. 

The compounds represented in Fig.~\ref{fig:ZF}b lie in the region of the phase diagram where the magnetic transition occurs simultaneously with the electronic and structural transitions. The temperature dependence of the frequency shows a discontinous onset at the ordering temperature for these compounds, corresponding to a first-order thermal phase transition. We also measured several compounds located further away from the QPT where the magnetic transition is split from the electronic and structural transitions. Representative ZF spectra for YNiO$_3$ taken at several temperatures spanning the thermal phase transition are shown in Fig.~\ref{fig:ZF}c, revealing a gradual increase in the precession frequency as the temperature is lowered. This is further illustrated in Fig.~\ref{fig:ZF}d for $RE=$Lu, Y, Eu, Sm, and Sm$_{0.75}$Nd$_{0.25}$, with the continuous onset of the oscillation frequency suggesting second-order-like behavior of the thermal phase transition. This is consistent with previous studies indicating a second-order magnetic transition for compounds with split electronic/structural and magnetic transitions, and first-order behavior when the transitions occur simultaneously~\cite{vobor;prb99}.

\begin{figure*}
	\includegraphics[width=150mm]{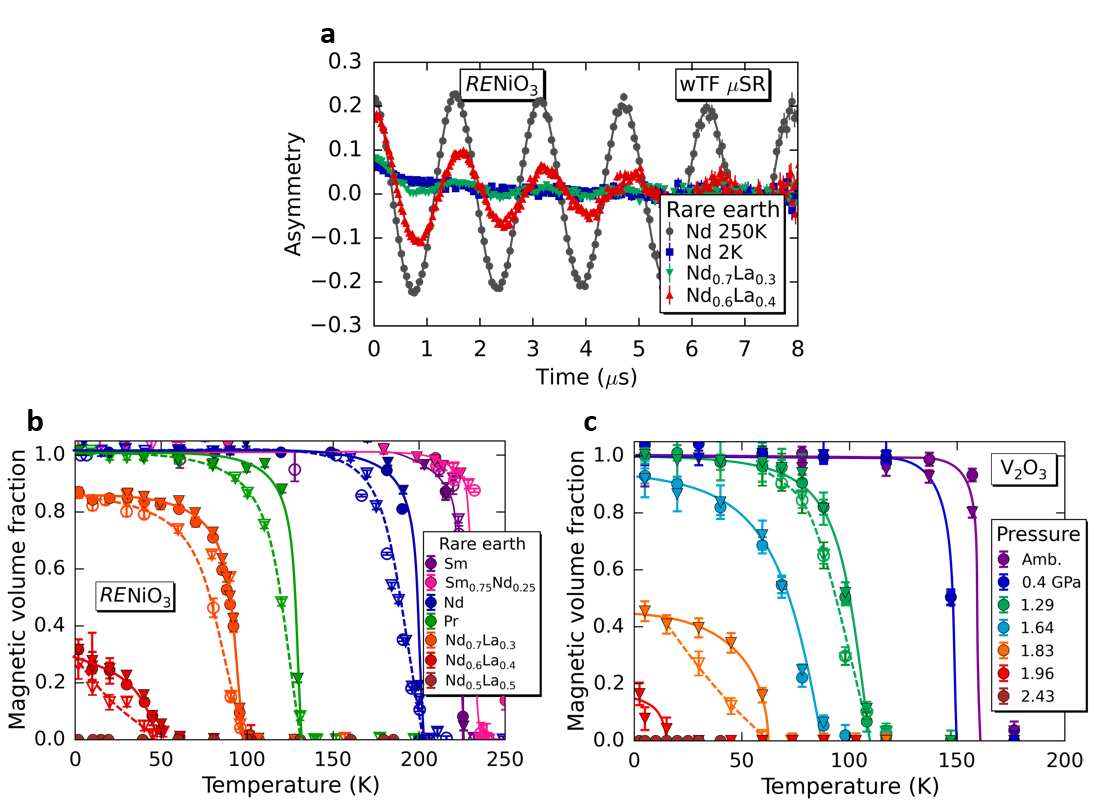}
	\caption{\label{fig:wTF} \textsl{Weak transverse field \musr\ experiments and magnetic volume fraction of \rno\ and \v2o3. (a) wTF time spectra for three compounds of \rno\ near the QPT at 2~K, with one spectrum measured at higher temperature (250~K) shown for comparison. The colored dots represent the data, the solid curves the fits described in the text. (b) Temperature dependence of the magnetic volume fraction in \rno\ derived from the fits. Circles (triangles) represent wTF (ZF) measurements, and filled (open) symbols represent warming (cooling) sequences. Solid and dashed curves are guides to the eye, with solid corresponding to cooling and dashed to warming. The compounds near the QPT have a significantly reduced ordered volume fraction at low temperature, indicating phase separation between magnetic and paramagnetic regions of the sample. Error bars were obtained by propagating the ESDs from the refined asymmetry values. (c) Magnetic volume fraction of \v2o3\ under different hydrostatic pressures. The symbols are the same as in (b). As with \rno, the ordered volume fraction is strongly reduced near the QPT.}}
	
\end{figure*}

The reduced ZF oscillation amplitudes observed in Fig.~\ref{fig:ZF}a for \nlnosf\ and \nlnost\ suggest that even at 2~K, these compounds are not magnetically ordered throughout the full volume fraction, rather there are some paramagnetic regions remaining. To verify this, we performed wTF experiments at several temperatures for each compound. The results are summarized in Fig.~\ref{fig:wTF}a-b. Panel (a) displays the time spectra for \nlnosf, \nlnost, and \nno\ at 2~K and \nno\ at 250~K. In wTF, the amplitude of the low-frequency oscillations is proportional to the non-magnetically-ordered volume fraction; thus, a spectrum with no oscillation (such as \nno\ at 2~K) corresponds to a fully ordered sample, while a spectrum with oscillation in the full asymmetry (such as \nno\ at 250~K) indicates a completely disordered sample. The spectra for \nlnost\ and \nlnosf\ show intermediate oscillation amplitudes, indicating that even at 2~K, these materials are only partially ordered. These compounds therefore exhibit intrinsic paramagnetic and antiferromagnetic phase separation. 

The temperature dependence of the magnetic volume fraction extracted from the wTF data for several compounds spanning the phase diagram is displayed by the circles in Fig.~\ref{fig:wTF}b, along with the results determined from the ZF experiments shown as triangles. \nlnost\ and \nlnosf, the compounds that are closest to the QPT, show a significantly reduced ordered volume fraction, verifying that their ground-state consists of phase-separated antiferromagnetic and paramagnetic regions. \pno\ and \nno\ also show phase separation over a broad temperature interval, although both are fully ordered at low temperature. The region of the phase diagram exhibiting phase separated behavior is indicated by the shaded area marked ``PS'' in Fig.~\ref{fig:phasediagram}. It is interesting to note that for the compounds exhibiting phase separation, the transition is ``stretched'' over a rather large temperature range. The hysteretic differences between measurements taken in cooling and warming sequences (represented by open and filled symbols, respectively) confirms the first-order nature of thermal phase transition for \nno, \pno, \nlnost, and \nlnosf. 

Importantly, the reduced magnetic volume fraction in \nlnost\ and \nlnosf\ cannot be attributed merely to an extrinsic effect of doping, since the doped compound \snno\ shows a rapid magnetic transition in the full volume fraction. Furthermore, neutron diffraction of the \nlnosf\ sample at 70~K indicates that it is single-phase (see Extended Data Fig. 2), excluding poor sample quality as a cause of this behavior. Therefore, we attribute the reduced volume fraction to proximity to the QPT. X-ray diffraction measurements on NdNiO$_3$ also confirm that the expected structural response at the MIT occurs with hysteresis over a similar temperature range as that of the magnetic hysteresis, verifying the correlation between the magnetic and structural phases (see Extended Data Fig. 1).




To investigate whether the volume-wise destruction of the AF Mott phase at the QPT is a feature unique to \rno\ or is perhaps more generic, we performed corresponding ZF and wTF experiments on pressure-tuned \v2o3. Hydrostatic pressure has the advantage of being a very clean tuning parameter, avoiding any possible complications of structural or chemical inhomogeneity often associated with chemical doping. The key results for \v2o3\ are identical to those for \rno: the AF state is destroyed by reduction of the ordered volume fraction while the moment size remains constant, with a significant region of ground-state phase separation near the QPT. The ZF \musr\ results for \v2o3\ are summarized in Fig.~\ref{fig:ZF}e-f. From the ZF spectra displayed in panel (e), it is clear that the oscillation frequencies at various pressures are virtually identical, while the oscillation amplitude shows a significant decrease as the critical pressure of $\sim$2.0~GPa is approached. As seen in panel (f), all ZF oscillation frequencies lie along the same two frequency bands, demonstrating that pressure does not affect the ordered moment size. The lower frequency band is in good agreement with an early \musr\ experiment~\cite{uemur;hfi84} on \v2o3. The ordered volume fraction obtained by both the ZF and wTF experiments is displayed in Fig.~\ref{fig:wTF}c, where we observe a gradual reduction in the low-temperature magnetic volume fraction beginning with the 1.64~GPa measurement. At 1.96~GPa, only about 15\% of the volume is magnetically ordered, and at 2.43~GPa, the magnetism is completely absent. The phase separation between AF and paramagnetic regions is reminiscent of that observed in the charge sector by spatially resolved infrared spectroscopy in pure and Cr-doped \v2o3.~\cite{mcleo;submitted15,lupi;nc10} The curious ``stretching'' of the magnetic transition observed in \rno\ is also seen in \v2o3\ for larger applied pressures, in agreement with an earlier high-pressure NMR study~\cite{takig;prl96}. This stretched transition is also somewhat similar to that observed in \v2o3\ nanocrystals~\cite{blago;prb10}, although surface area effects are likely the cause in that case. These results demonstrate that the AF phases in \rno\ and \v2o3\ have identical behavior in all key aspects near the QPT.

The two main results of these experiments are 1) that the saturated magnetic moment size is constant across the ordered region of the phase diagram before being abruptly destroyed at the QPT; and 2) that the magnetic volume fraction is heavily reduced near the QPT. We observed identical behavior for \rno\ and pressure-tuned \v2o3, both canonical bandwidth-controlled Mott insulators. These results unambiguously demonstrate that the QPT in these materials proceeds in a distinctly first-order fashion. The striking similarity between both of these canonical systems, along with experimental indications of structural phase separation in the archetypal Mott system NiS$_2$ tuned by pressure~\cite{feng;prb11}, indicates that this is a generic feature of the pressure-driven Mott QPT, supporting previous theoretical work~\cite{watan;jpsj04,misaw;prb07,imada;prb05,chitr;prl99,park;prb14}. Further theoretical considerations regarding the surprisingly large region of parameter space exhibiting phase coexistence and additional details of this first-order QPT are provided in the Supplementary Discussion.

We note that similar phase separation between paramagnetic and magnetically ordered regions was observed by \musr~\cite{uemur;np07} and nuclear magnetic resonance~\cite{yu;prl04} in the destruction of magnetic order in the itinerant helimagnet MnSi through pressure tuning, and by \musr~\cite{uemur;np07} in the metallic ferromagnet (Sr,Ca)RuO$_3$ through (Sr,Ca) chemical substitution. Signatures of first-order magnetic quantum evolution have also been observed in heavy fermion systems~\cite{uemur;np07} and many unconventional superconductors,~\cite{uemur;np07,uemur;nmat09} accompanied by an inelastic resonance mode in the superconducting state.~\cite{uemur;nmat09,uemur;jpcm04} The magnetic resonance mode can be viewed as a soft mode toward the ``parent''/``competing'' magnetic state appearing due to the closeness of free energies of the magnetically ordered and superconducting states across the first-order transition.~\cite{uemur;nmat09} Such a soft mode may also exist in the paramagnetic metallic state of non-superconducting Mott transition systems near the QPT as a signature of the imminent AF-insulating electronic structure appearing in the dynamic and inelastic spin/charge correlations. The pseudogap observed in tunneling experiments~\cite{allen;arxiv14} on paramagnetic metallic LaNiO$_3$ may be viewed as a charge soft mode in a magnetically disordered metallic state adjacent to the Mott insulator NdNiO$_3$. Further studies of antiferromagnetic Mott insulators and other quantum magnetic systems will be useful to elucidate generic and system-specific roles of first-order behavior in quantum phase evolution.

\textbf{Methods}

\textbf{Specimen preparation.}
The \rno\ perovskites were prepared in polycrystalline form as follows: analytical grade $R_2$O$_3$ ($R$=La,Pr,Nd,Sm) and Ni(NO$_3$)$_2 \cdot$6H$_2$O were solved in citric acid with some droplets of nitric acid; the mixture of citrate and nitrate solutions was slowly evaporated, leading to organic resins, which were dried and decomposed by slowly heating up to 800$^{\circ}$C in air, for 12 hours. This treatment gave rise to highly reactive precursor materials, amorphous to X-ray diffraction. The precursor powders were treated at 900$^{\circ}$C under 200~bar of O$_2$ pressure for 12 hours in a Morris Research furnace. Then the samples were slowly cooled down ( 2$^{\circ}$C min$^{-1}$) to room temperature. Finally, the samples were pelletized and re-treated at 900$^{\circ}$C under O$_2$ pressure for 12 hours to give 6~mm disks suitable for the \musr\ experiments. Another set of (Nd,La)NiO$_3$ samples for additional x-ray measurement was prepared from a stoichiometric mixture of $RE_2$O$_3$(99.9\%), NiO(99.9\%) and KClO$_4$ (99\%, 100\% overweight). The mixtures were placed in a Au capsule and treated at 6GPa in a cubic-anvil-type high pressure apparatus at 1100$^{\circ}$C for 30 min before being quenched to room temperature and subsequently releasing the pressure. After removing the capsule, the sample was crushed and washed in distilled water to dissolve the KCl and obtain the pure $RE$NiO$_3$ sample. The resulting $RE$NiO$_3$ purified powder was heated in an evacuated oven at 150$^{\circ}$C for 30 h to evaporate any remaining water. The large polycrystalline specimen of \v2o3\ was synthesized via reduction of high-purity V$_2$O$_5$ in 5\% H$_2$/Ar gas at 900~$^{\circ}$C for 48 hours, followed by cooling at 100~$^{\circ}$C/h. 

\textbf{\musr\ experiments.}
The \musr\ experiments on \rno\ were conducted at the Centre for Molecular and Materials Science at TRIUMF in Vancouver, Canada using the LAMPF spectrometer, and the pressure-dependent \musr\ experiments on \v2o3\ were conducted at the Paul Scherrer Institute in Villigen, Switzerland using the GPD instrument. In both cases, a gas-flow cryostat was used, providing access to temperatures from 2~K to room temperature. Hydrostatic pressure was applied to \v2o3\ by immersing the sample in Daphne oil in a double-walled cylindrical piston cell made of MP35N material. The pressure was calibrated by observing the change in the superconducting transition temperature of a small indium plate placed in the oil with the sample. The uncertainty in the measured pressure was less than 0.05~GPa. 

The \musr\ spectra were analyzed using the least-squares minimization routines in the MusrFit software package~\cite{suter;physproc12}. The wTF asymmetry spectra were modeled by the function 
\begin{equation}\label{wTFfunc}
A(t)=a_s \exp(-\Lambda t) \cos(\omega t + \phi),
\end{equation}
where $t$ is time after muon implantation, $A(t)$ is the time-dependent asymmetry, $a_s$ is the amplitude of the oscillating component, $\Lambda$ is an exponential damping rate due to paramagnetic spin fluctuations and/or nuclear dipolar moments, $\omega$ is the Larmor precession frequency set by the strength of the transverse magnetic field, and $\phi$ is a phase offset. In the case of \v2o3, a second slowly-damped oscillating component was included to account for muons stopping in the pressure cell. The ``zero'' for $A(t)$ was allowed to vary for each temperature to deal with the asymmetry baseline shift known to occur for magnetically ordered samples. From these refinements, the magnetically ordered volume fraction at each temperature $T$ was estimated as $1-\frac{a_s (T)}{a_s (T_{\mathrm{max}})}$, where $a_s (T_{\mathrm{max}})$ is the amplitude in the paramagnetic phase at high temperature.

The ZF asymmetry spectra were modeled by the function
\begin{equation}\label{ZFfunc}
A(t)=\sum_{i=1}^{n}a^{\mathrm{osc}}_i \exp(-\Lambda^{\mathrm{T}}_{i} t) \cos(2\pi\nu_i t)+a^{\mathrm{non}}\exp(-\Lambda^{\mathrm{L}} t),
\end{equation}
where the $a^{\mathrm{osc}}_i$ are the amplitudes of the precessing components, the $\Lambda^{\mathrm{T}}_{i}$ are the transverse damping rates due to a distribution of field strengths at the muon site(s), the $\nu_i$ are the precession frequencies, $a^{\mathrm{non}}$ is the asymmetry of the non-oscillating component due to a combination of paramagnetic regions and the ``1/3'' non-oscillating component arising from the orientationally averaged magnetic regions of the sample, and $\Lambda^{\mathrm{L}}$ is the longitudinal relaxation rate (also known as 1/T$_1$ when it applies to the 1/3 component from magnetically ordered regions). In all cases except for $RE=\mathrm{Nd}$, there are two oscillating components at the lowest temperatures, with the lower-frequency component having roughly double the amplitude of the higher-frequency component. This may be understood as arising from two magnetically inequivalent muon stopping sites. For $RE=\mathrm{Nd}$, a third frequency can be resolved at the lowest temperatures, perhaps due to ordering of the Nd moments. For the \v2o3\ experiments, a damped Kubo-Toyabe component with fixed amplitude and field width was also included to capture the contribution from muons stopping in the pressure cell. The total initial asymmetry was approximately 0.258, with $0.101/0.258\simeq 40\%$ of the signal arising from muons stopping in the sample. 

The precession frequencies were determined from refinements performed over the first $\mu$s of the asymmetry time spectra. Additional refinements over a larger time window (0-8$\mu$s) and a larger time-channel binning size were performed to more accurately determine $a^{\mathrm{non}}$ and $\Lambda^{\mathrm{L}}$. For a magnetically ordered sample with domains distributed isotropically over all solid angles, 2/3 of the asymmetry will exhibit oscillations and 1/3 will not. This is because magnetic field components along the Cartesian direction defined by the initial muon spin do not contribute to the precession of the muon spin, while the other two Cartesian directions do. In \rno\ and \v2o3, as in many other magnetic materials, the ``2/3'' oscillating component is strongly damped, and is manifest only as ``missing asymmetry'' when the spectrum is viewed over a large time window with relatively large time-channel binning. The total asymmetry arising from magnetically ordered regions of the sample can therefore be estimated as 3/2 multiplied by the missing asymmetry, with the magnetic volume fraction then given by $3/2 \times \frac{a^{\mathrm{non}}(T_{\mathrm{max}})-a^{\mathrm{non}}(T)}{a^{\mathrm{non}}(T_{\mathrm{max}})}$. For \rno, the magnetic volume fraction determined from ZF data in this way agreed quite well with the volume fraction determined by the wTF experiments. For \v2o3, this resulted in a slight overestimation of the magnetic volume fraction. Therefore, we multiplied by 4/3 instead of 3/2 (corresponding to a 1/4 non-oscillating component rather than 1/3), which then gave close agreement with the wTF results. Such a deviation from the ideal 2/3 versus 1/3 division of the asymmetry is not uncommon and simply reflects an imperfect orientational average of the magnetic domains.

\textbf{X-ray and neutron diffraction measurements.}
X-ray diffraction measurements of NdNiO$_3$ were performed at NSLS-II at Brookhaven National Laboratory on the XPD instrument~\cite{wang;aplmat15,jense;iucrj15}. The temperature was controlled with a gas cryostream. Rietveld refinements~\cite{rietv;jac69} of the diffraction patterns were performed using the Full Prof software suite~\cite{rodri;pb93}, and pair distribution functions (PDFs) were generated and modelled using the xPDFsuite program~\cite{yang;arxiv15}. The orthorhombic Pbnm structure was used to model the data at all temperatures. This was sufficient to verify the presence of the expected structural change over the same temperature region as the magnetic transition, even though the true low-temperature structure has monoclinic symmetry. Supplementary Figure 1 displays the temperature dependence of the unit cell volume in panel (a) and the isotropic atomic displacement parameter for Ni in panel (b), both obtained from PDF modeling. In both cases, a deviation from the high-temperature trend is observed at 200~K, with substantial differences between the warming and cooling sequences over a similar temperature range as the magnetic volume fraction hysteresis observed by \musr. The results of the Rietveld refinements were very similar. These observations are consistent with previous structural studies of NdNiO$_3$.

Time-of-flight neutron diffraction measurements of Nd$_{0.6}$La$_{0.4}$NiO$_3$ were performed at the Spallation Neutron Source of Oak Ridge National Laboratory on the NOMAD instrument. Rietveld refinements were performed using GSAS~\cite{larso;unpub94} on the EXPgui platform~\cite{toby;jac01}. Supplementary Figure 2 displays the results of refining the Pbnm model against the measured diffraction pattern at 70~K, which is above the MIT in this compound. The measured pattern is shown by blue crosses, the calculated pattern in red, and the difference curve in green. The fit is of good quality, ruling out the presence of any significant impurity phase in this compound. 

\textbf{Acknowledgements}
The authors acknowledge helpful discussions with Dietrich Belitz, Andy Millis, and Dimitri Basov. Work at Columbia University was supported by the U.S. National Science Foundation (NSF) via Grant DMREF DMR-1436095, NSF Grant No. DMR-1105961, and the NSF PIRE program through Grant No. OISE-0968226, with additional support from the Japan Atomic Energy Agency Reimei Project and the Friends of Todai Foundation. BAF acknowledges support from the NSF GRFP under Grant No. DGE-11-44155. SJLB acknowledges support from the U.S. Department of Energy, Office of Science, Office of Basic Energy Sciences (DOE-BES) under contract No. DE-SC00112704. Work at McMaster was supported by  NSERC. JAA acknowledges financial support from MINECO (SPAIN) through the project MAT2013-41099-R. ZG acknowledges support by the Swiss National Science Foundation. Work at Kyoto University was supported by the Grant-in-Aid for Scientific Research (A) (No. 25248016). Work at the Chinese Academy of Sciences was supported by the Chinese NSF and MOst. Work at Zhejiang was supported by the Chinese NSF (No.11274268 and 11574265). GK acknowledges support from the U.S. NSF through Grant DMREF DMR-1435918. MI thanks financial support from a Grant-in-Aid for Scientific Research (No. 22104010) from MEXT, Japan, and by MEXT HPCI Strategic Programs for Innovative Research (SPIRE) (under the grant number hp130007, hp140215 and hp150211) and Computational Materials Science Initiative (CMSI). Use of the National Synchrotron Light Source II, Brookhaven National Laboratory, was supported by DOE-BES under contract No. DE-SC0012704. Use of the Spallation Neutron Source, Oak Ridge National Laboratory, was sponsored by the Scientific User Facilities Division, Office of Basic Energy Science, U.S. DOE.

\textbf{Author contributions}
This project was proposed and coordinated by YJU. \musr\ data were taken by BAF, LL, SCC, TJSM, AMH, MNW, YC, GL, BC, CD, SG, FN, TUI, WH, SS, AF, and YJU for \rno\, and by BAF, SCC, ZG, RK, EM, and YJU for \v2o3. The \musr\ data analysis was carried out by BAF and YJU. The neutron and x-ray scattering studies were performed by BAF and SJLB. JA, CJ, and WML provided the \rno\ samples. HK and TM provided the \v2o3\ sample. MI and GK provided theoretical considerations. The manuscript was drafted by BAF and YJU, and has been completed with input from all authors. This work is part of the PhD thesis project of BAF at Columbia University.

\textbf{Additional information}
Methods and extended data can be found in the online version of this paper.

\textbf{Competing financial interests}

The authors declare no competing financial interests.

\end{document}


\preprint{}

\title{
Supplementary Information: Volume-wise destruction of the antiferromagnetic Mott insulating state through quantum tuning
}

\author{Benjamin A. Frandsen}
\affiliation{%
 Department of Physics, Columbia University, New York, NY 10027, USA.
}%

\author{Simon J. L. Billinge}
\affiliation{%
	Department of Applied Physics and Applied Mathematics, Columbia University, New York, NY 10027, USA.
}%
\affiliation{%
	Condensed Matter Physics and Materials Science Department, Brookhaven
	National Laboratory, Upton, New York 11973, USA.
}%

\author{Gabriel Kotliar}
\affiliation{ %
Department of Physics \& Astronomy, Rutgers University, Piscataway, New Jersey 08854-8019, USA.
} %

\author{Masatoshi Imada}
\affiliation{ %
Department of Applied Physics, University of Tokyo, 7-3-1 Hongo, Tokyo, 113-8656, Japan
} %

\author{Yasutomo J. Uemura}
\email{tomo@lorentz.phys.columbia.edu}
\affiliation{%
 Department of Physics, Columbia University, New York, NY 10027, USA.
}%

\date{\today}

\maketitle

\begin{figure*}
	\includegraphics[width=150mm]{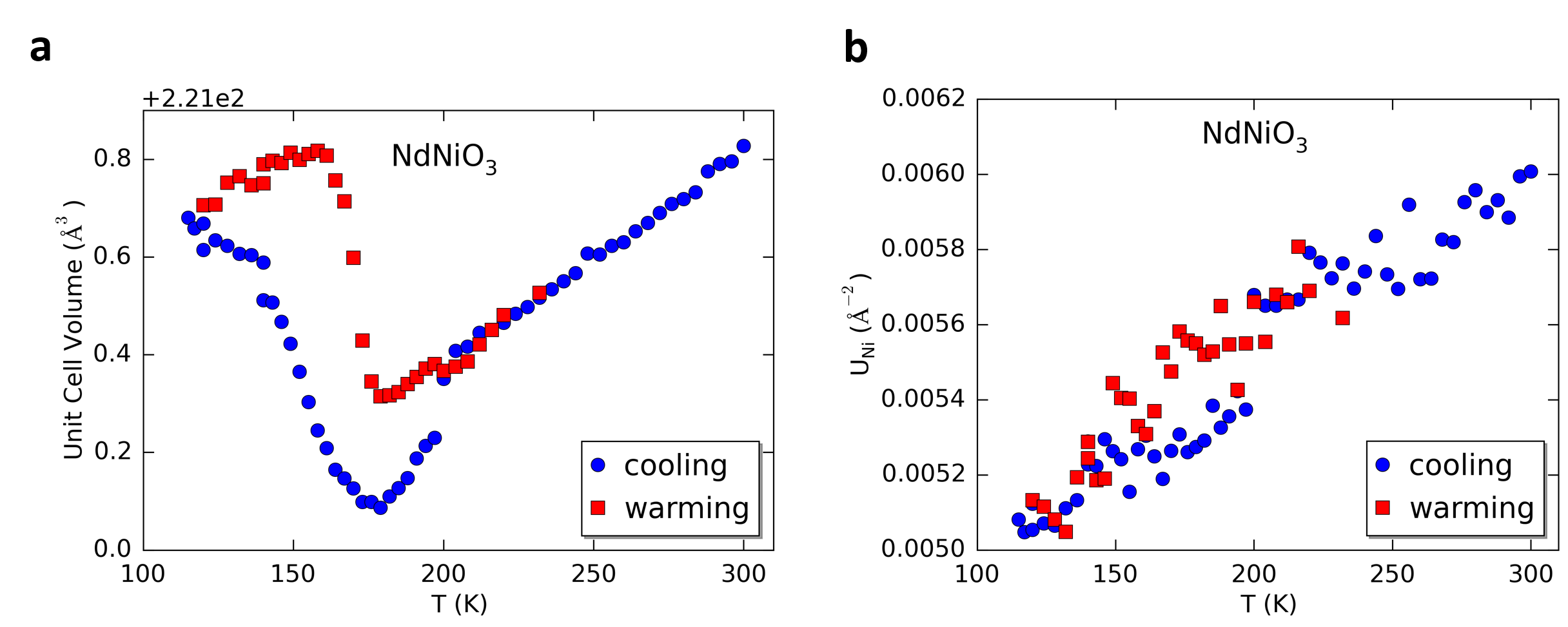}
	\caption{\label{fig:xray} \textbf{(Supplementary Figure 1) Selected results of structural refinements of temperature-dependent x-ray pair distribution function (PDF) measurements of NdNiO$_3$.} (a) Unit cell volume obtain from refined lattice parameters. A clear response is observed at 200~K with significant hysteresis over a large temperature region, consistent with the \musr\ results. (b) Isotropic atomic displacement parameter $U$ for Ni, again showing a hysteretic response below 200~K.}
	
\end{figure*}

\begin{figure*}
	\includegraphics[width=120mm]{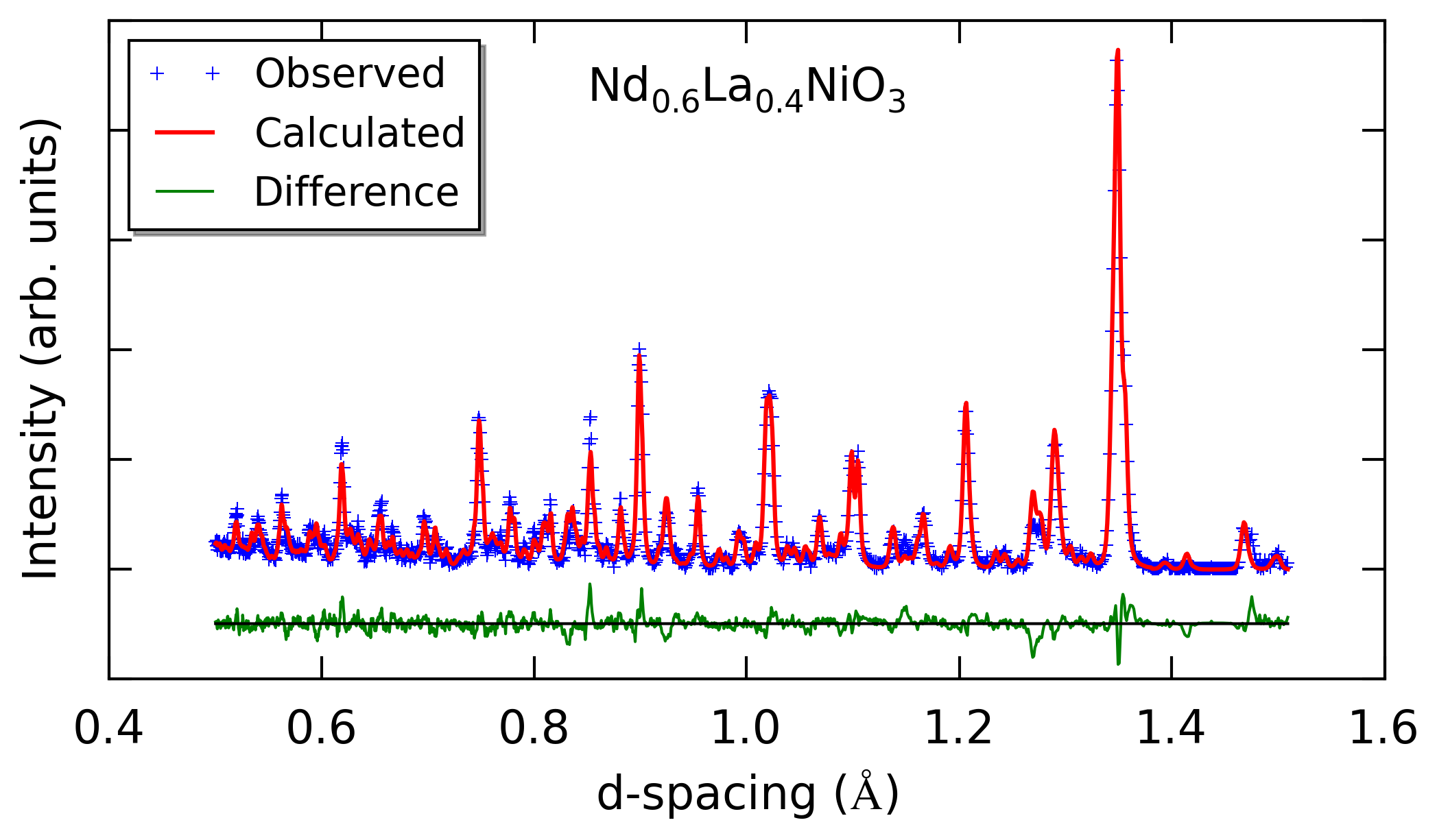}
	\caption{\label{fig:neutron} \textbf{(Supplementary Figure 2) Neutron diffraction measurement of Nd$_{0.6}$La$_{0.4}$NiO$_3$ at 70~K.} The blue crosses represent the measured intensity, the red curve is the calculated pattern, and the green curve is the difference.}
	
\end{figure*}

\textbf{Supplementary Discussion: Theoretical considerations regarding the metal-insulator transition}

In an ideal system, the pressure-driven MIT should occur at one specific pressure for any given temperature. However, the disorder resulting from the use of chemical pressure and the finite time scales involved in the experimentation necessarily introduce microscopic inhomogeneities and non-equilibrium phenomena, which are manifest in a finite interval of pressures where the two phases coexist. Additionally, the assumption of constant pressure may be unrealistic at the microscopic level due to local variations in unit cell structure and/or size, causing microscopic strains near domain boundaries, for example. This further leads to inhomogeneous nucleation during the phase transition. Recent theoretical work of Yee and Balents~\cite{yee;prx15} also showed phase separation between antiferromagnetic Mott insulating and paramagnetic metallic states via charge doping, but this may not be directly applicable to the present work due to differences in the tuning methods and the underlying physical assumptions.

The universality brought to light in this paper suggests that while magnetism and structural changes accompany the Mott transition, these effects are not likely to be essential to its existence, since they are present with various strengths in different compounds. Earlier theoretical results~\cite{watan;jpsj04,misaw;prb07,imada;prb05,chitr;prl99} and the experimental results presented in this work suggest that the strength of the first-order transition can be varied, and decreased with increasing quantum fluctuations to approach an interesting marginal point which separates the first order regime from a regime where the Mott transition is continuous. The marginal point may offer novel physics associated with the quantum criticality distinct from the conventional one driven by the spontaneous symmetry breaking. The pressure-driven metal-insulator coexistence region revealed in the present paper provides us with the possibility of unique quantum fluctuations reminiscent of, but distinct from, the inhomogeneity extensively studied in the filling controlled case such as the high-T$_c$ cuprates and the colossal magnetoresistive perovskite manganites. The generality  of the first-order nature of the Mott transition across different series of materials calls for the determination of its strength. Detailed studies of transport  coefficients~\cite{husma;s96}, optical conductivities, and sound velocity measurements coupled with system specific calculations which go beyond model studies will be  needed to answer this question.